\documentclass[a4paper,11pt]{article}
\usepackage[sort&compress,numbers]{natbib}
\usepackage[utf8]{inputenc}
\usepackage{amsmath,amssymb,mathtools}
\usepackage{graphicx}
\usepackage{xcolor}
\usepackage{multirow}
\usepackage{babel}
\usepackage{xspace}
\usepackage{enumerate}
\usepackage{caption}
\usepackage{subcaption}
\usepackage{soul}
\usepackage{soul}
\usepackage{float} % Con el flag H fija la figura en esa posicion en el texto
\usepackage{comment}
\usepackage{longtable}

\def\beq{\begin{equation}}
\def\eeq{\end{equation}}
\def\bea{\begin{align}}
\def\eea{\end{align}}

\def\Eq#1{Eq.~(\ref{#1})}

\def\ra{\rangle}
\def\la{\langle}

\def\ket#1{|#1\ra}

\usepackage{pos}

\title{Quantum jet clustering with LHC simulated data}
\author*{Jorge J. Martínez de Lejarza}
\author{Leandro Cieri}
\author{Germán Rodrigo¹}

\affiliation[]{
  Instituto de F\'{\i}sica Corpuscular, Universitat de Val\`encia - Consejo Superior de Investigaciones Cient\'{\i}ficas, Parc Cient\'{\i}fic, E-46980 Paterna, Valencia, Spain}

\note{Disclaimer: GR currently on leave at the European Research Council Executive Agency, European Commission, BE-1049 Brussels, Belgium. The views expressed are purely those of the writer and may not in any circumstances be regarded as stating an official position of the European Commission.}

\emailAdd{Jorge.M.Lejarza@ific.uv.es}
\emailAdd{lcieri@ific.uv.es}
\emailAdd{german.rodrigo@csic.es}

\abstract{We study the case where quantum computing could improve jet clustering by considering two new quantum algorithms that might speed up classical jet clustering algorithms. The first one is a quantum subroutine to compute a Minkowski-based distance between two data points, while the second one consists of a quantum circuit to track the rough maximum into a list of unsorted data. When one or both algorithms are implemented in classical versions of well-known clustering algorithms (\texttt{K-means}, Affinity Propagation and $k_T$-jet) we obtain efficiencies comparable to those of their classical counterparts. Furthermore, in the first two algorithms, an exponential speed up in dimensionality and data length can be achieved when applying the distance or the maximum search algorithm. In the $k_T$ algorithm, a quantum version of the same order as \texttt{FastJet} is achieved.}

\FullConference{%
  41st International Conference on High Energy physics - ICHEP2022\\
  6-13 July, 2022\\
  Bologna, Italy
}

\begin{document}
\maketitle

\section{Introduction}
Quantum computing devices, based on the laws of quantum mechanics, offer the possibility to efficiently solve specific problems that become very complex or even unreachable for classical computers since they scale either exponentially or super-polynomially.
Very recently, quantum algorithms have started to be applied in solving problems which appear in high-energy particle physics. Recent applications of quantum algorithms to HEP cover diverse subareas such as jet clustering ~\cite{Wei:2019rqy,Pires:2021fka,Delgado:2022snu,deLejarza:2022bwc} or  multi-loop Feynman integrals~\cite{Ramirez-Uribe:2021ubp}. For a recent review on the applications of quantum computing to data analysis in HEP we refer the reader to Ref.~\cite{Delgado:2022tpc}.

In the present work, which is based on Ref.~\cite{deLejarza:2022bwc}, we address the problem of clustering and jet reconstruction from collider data, which is a nontrivial and computationally expensive task, as it often involves performing optimizations over potentially large numbers of final-state particles. In particular, we consider the possibility of using quantum algorithms to improve the velocity in jet identification. Here we focus on three well-known classical algorithms:  \texttt{K-means} clustering~\cite{macqueen1967some}, Affinity Propagation (\texttt{AP})~\cite{Frey2007ClusteringBP} and $k_T$-jet clustering in all its variants~\cite{Catani:1991hj,Catani:1993hr,Ellis:1993tq,Cacciari:2008gp}. We propose the corresponding quantum versions of the precedents algorithms: quantum \texttt{K-means} clustering, quantum \texttt{AP}-algorithm and quantum $k_T$-based algorithms.

These quantum versions made use of two new quantum subroutines proposed in this work. 
The first one serves to compute distances satisfying Minkowski metric, whereas the second one consists of a quantum circuit to track the maximum into a list of unsorted data.

%%%%%%%%%%%%%%%%%%%%%%%%%%%%%%%%%%%%%%%%%%%%%%%%%%%%%
%%%%%%%%%%%%%%%%%%%%%%%%%%%%%%%%%%%%%%%%%%%%%%%%%%%%%
\section{Quantum distance in Minkowski space}
\label{sec:qdis}
In quantum computing, it is essential to have the ability to measure quantum entanglement between two states, as in many cases it determines the possibility of obtaining a quantum advantage.  We rely on the \textit{SwapTest} method~\cite{Buhrman:2001} to quantify the similarity of two states.

\begin{figure}[th!]
    \centering
    \vspace{-.5cm}
    \includegraphics[width=0.4\textwidth]{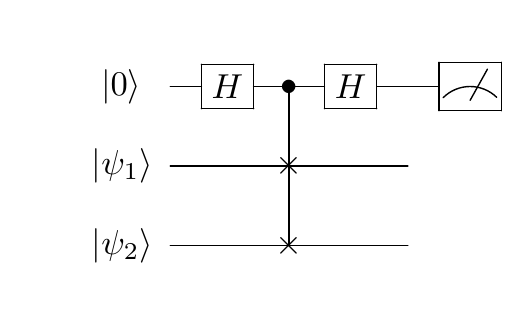} \vspace{-.5cm}
    \caption{Quantum circuit \textit{SwapTest}.}  
    \label{fig:qsmassinvariant}
\end{figure}

As it can be seen in Fig. \ref{fig:qsmassinvariant}, the \textit{SwapTest} procedure makes use of one auxiliary qubit to get the inner product between two states:
\beq
P(\ket{0}) =
\frac{1}{2}+\frac{1}{2} \left|\la \psi_1| \psi_2 \ra \right|^2~.
\label{eq:p0swaptestsimple}
\eeq
The \Eq{eq:p0swaptestsimple} displays that a distance between two quantum vectors can be obtained just measuring an ancillary qubit.

%\label{subsec:qinvmass}
%%%%%%%%%%%%%%%%%%%%%%%%%%%%%%%%%%%%%%%%%%%%%%%%%%%%%%%%%%%%%%%%%%%
%%%%%%%%%%%%%%%%%%%%%%%%%%%%%%%%%%%%%%%%%%%%%%%%%%%%%%%%%%%%%%%%%%%

Vectors in high-energy physics are defined in a
four-dimensional space-time with Minkowski metric. They have 
the form $x_i = (x_{i,0}, {\bf x}_i)$, where $x_{i,0}$ is the 
temporal component and ${\bf x}_i$ represent the three spatial components. In the following, we assume that the dimension of the space-time is $d$, where $d-1$ is the number of spatial components. We shall define the analogue of the Euclidean classical distance in the Minkowski space corresponding to the invariant sum squared $s_{ij}^{\rm (C)}$, which is commonly called invariant mass squared when vectors are particle four-momenta,
\beq
s_{ij}^{\rm (C)} = (x_{0,i}+x_{0,j})^2 
- |{\bf x}_i + {\bf x}_j|^2~. 
\eeq

This quantity, which is Lorentz invariant, can be used as test distance to measure similarity between particle momenta. It is necessary to apply twice the \textit{SwapTest} subroutine for computing the Minkowski-type distance through a quantum algorithm. Once for the spatial and once for the temporal component.

The spatial distance is computed using the subsidiary states: \beq
\ket{\psi_1} =\frac{1}{\sqrt{2}} \left( \ket{0, x_i} + \ket{1, x_j} \right)~, \qquad
\ket{\psi_2} =\frac{1}{\sqrt{Z_{ij}}} \left( |{\bf x}_i| \ket{0} -|{\bf x}_j| \ket{1} \right)~,
\label{eq:varphi}
\eeq
where $Z_{ij}=|{\bf x}_i|^2+|{\bf x}_j|^2$ 
is a normalization factor and $\ket{0}$ and $\ket{1}$ are the states of an ancillary qubit.
Whereas the temporal distance is computed as a result of the overlap of the following states:
\begin{equation}
\ket{\varphi_1} = H \ket{0} = \frac{1}{\sqrt{2}} \left( \ket{0} + \ket{1} \right)~, \qquad 
\ket{\varphi_2} =\frac{1}{\sqrt{Z_0}} \left(x_{0,i} \ket{0} +x_{0,j} \ket{1} \right)~,
\label{eq:minkstates}
\end{equation}
where $Z_{0}=x_{0,i}^2+x_{0,j}^2$. Then, applying the \textit{SwapTest} to the states $\ket{\psi_1}$, $\ket{\psi_2}$ and $\ket{\varphi_1}$, $\ket{\varphi_2}$ one gets the relations:

\begin{equation}
P(|0\rangle|_{spatial})=\frac{1}{2}+\frac{1}{2} |\langle \psi_1| \psi_2 \rangle|^2~, \qquad 
P(|0\rangle|_{time})=\frac{1}{2}+\frac{1}{2} |\langle \varphi_1| \varphi_2 \rangle|^2 \ ,
\label{eq:p0time}
\end{equation}
where the overlaps $|\langle \psi_1| \psi_2 \rangle|^2$ and $|\langle \varphi_1| \varphi_2 \rangle|^2$ are given by
\begin{equation}
|\langle \psi_1| \psi_2 \rangle|^2 =\frac{1}{2Z_{ij}}|{\bf x}_i-{\bf x}_j|^2~, \qquad 
|\langle \varphi_1| \varphi_2 \rangle|^2 = \frac{1}{2Z_0}(x_{0,i} +x_{0,j})^2.
\label{eq:phioverlap}
\end{equation}
At this point, the quantum version of the invariant sum squared follows from the combination of results from \Eq{eq:p0time} and \Eq{eq:phioverlap}:
\beq
s_{ij}^{\rm (Q)} =2\big( 
Z_0(2P(|0\rangle|_{time})-1)-Z_{ij}(2P(|0\rangle|_{spatial})-1)\big).
\label{eq:qdistancemink}
\eeq

%%%%%%%%%%%%%%%%%%%%%%%%%%%%%%%%%%%%%%%%%%%%%%%%%%%
%%%%%%%%%%%%%%%%%%%%%%%%%%%%%%%%%%%%%%%%%%%%%%%%%%%
\section{Quantum maximum search}
\label{sec:qsearch}
Finding a particular element belonging to a dataset is a recurring problem in data analysis. This is a computationally very expensive task. However, quantum computing offers suitable tools to solve data query in a shorter computational time. In particular,  it is well known the quadratic speed up exhibited by Grover's algorithm~\cite{Grover:1997fa}.
In this work, we present a considerably simpler algorithm that is used exclusively to find the rough maximum in a list of values. 
%This algorithm, although very elementary, is sufficiently accurate for the applications that we will present in Sections~\ref{subsec:qkmeans} and~\ref{subsec:qktalgorithm}. To our knowledge, it is the first time presented in the literature.

Let $L[0, \ldots ,N-1]$ be an unsorted list of $N$ items. Solving the maximum searching problem is to find the index $y$ such that $L\left[y\right]$ is the maximum. For clustering algorithms, $L\left[y\right]$ represents the inverse of the distance. The quantum algorithm to solve that problem using amplitude encoding proceeds in two steps:

%\begin{enumerate}
    1.  The list of $N$ elements is encoded into a $\log_2(N)$ qubits state as follows:
    \begin{equation}
    \ket{\Psi} = \frac{1}{\sqrt{L_{sum}}}\sum_{j=0}^{N-1}L\left[j\right] \, \ket{j}~,
        \label{eq:amplitude encoding}
    \end{equation}
    where $L_{sum}=\sum_{j=0}^{N-1}L[j]^2 $ is a normalization constant.
    %This amplitude encoding is achieved using qRAM.
    
    2.  The final state is measured. 
    
\noindent The procedure is rerun several times to reduce the statistical uncertainty. Once done, the most repeated state gives us the rough maximum.
%\end{enumerate}
\section{Quantum clustering algorithms}\label{sec:qalgorithms}
\subsection{K means}
\texttt{K-means} is an unsupervised machine learning algorithm that classifies the elements of a dataset into $K$ groups called clusters~\cite{macqueen1967some}. % The data points within each cluster have to be as similar (near) as possible whereas the clusters themselves have to be as different (far) as possible from each other. 
The input for this algorithm is a set of $N$ data points or vectors in $d$ dimensions as well as the number of clusters $K$, with $K\le N$, and its output is a set of $K$ centroids.
The main core of this algorithm is computing distances among data points and then assigning the data points to their nearest centroid. That might be speed up using the quantum algorithms explained in Sec.~\ref{sec:qdis} and \ref{sec:qsearch} respectively assuming the data has been loaded from a quantum Random Access Memory (qRAM). In particular, we would obtain a speed up from $\mathcal{O}(NKd)$ in the classical version to $\mathcal{O}(N\log K \log(d-1))$ in our quantum version, where the speed-up in the dimensionality $d$ would be achieved when the quantum distance subroutine is applied and the speed-up number of clusters $K$ would be achieved when the minimum is obtained by the quantum maximum search algorithm.

\subsection{Affinity Propagation}
The Affinity Propagation algorithm is an unsupervised machine learning algorithm that, in contrast to \texttt{K-means}, does not need the number of clusters $K$ to be defined beforehand \cite{Frey2007ClusteringBP}. Its main ideas are that all data points are consider as exemplars (clusters) and what is called ``passing messages between data points''. That is achieved calculating the responsibility matrix $R(i,k)$, that measures how well-suited $k$ is to serve as an exemplar for $i$, and the availability matrix, that gauges how appropriate it would be for $i$ to choose $k$ as its exemplar. These matrices are computed based on the so-called similarity matrix.
This matrix includes a distance as a metric to quantify the similarity among data points. This distance will be, in our study, the invariant sum squared shown in \Eq{eq:amplitude encoding} and it might be computed quantumly. In fact, one would obtain a speed-up from $\mathcal{O}(N^2Td)$ to $\mathcal{O}(N^2T\log(d-1))$, where $N$ is the number of data points, $T$ is the iterations of the algorithm and $d$ the dimensionality.

\subsection{$k_T$ jet algorithm}
\noindent The inclusive variant of the generalised $k_T$-jet algorithm is formulated as follows:

%\begin{enumerate}
    1. For each pair of partons $i$, $j$ the following distance is computed:
    \begin{equation}
        d_{ij}=\mathrm{min}  (p_{T,i}^{2p},p_{T,j}^{2p}) \Delta R_{ij}^2/R^2,
        \label{eq:kt}
    \end{equation}
    with $\Delta R_{ij}^2=(y_i-y_j)^2+(\phi_i-\phi_j)^2$, where $p_{T,i}$, $y_i$ and $\phi_i$ are the transverse momentum (with respect to the beam direction), rapidity and azimuth of particle $i$. $R$ is a jet-radius parameter usually taken of order 1. For each particle $i$ the beam distance is $d_{iB}=p_{T,i}^{2p}$.
    
    2. Find the minimum $d_{min}$ amongst all the distances $d_{ij}$, $d_{iB}$. If $d_{min}$ is a $d_{ij}$, the particles $i$ and $j$ are merged into a single particle summing their four-momenta; if $d_{min}$ is a $d_{iB}$ then the particle $i$ is declared as a final jet and it is removed from the list.
    
    3. Repeat from step 1 until there are no particles left.
%\end{enumerate}

\noindent The step 2 of the above flow chart might be speed up using the quantum search algorithm presented in Sec.~\ref{sec:qsearch}. Then, the quantum $k_T$-jet algorithm would be $\mathcal{O}(N \log N)$ which outperforms the $\mathcal{O}(N^2)$ version of the $k_T$-jet algorithm and is of the same order as the \texttt{FastJet} version.

\section{Quantum simulations}
In Table \ref{tab:eff}, the quantum algorithms described above are tested with a simulated physical $n$-particle event produced at the LHC. The efficiency $\varepsilon_c$, defined as the quotient of the number of particles clustered in the same way as their classical counterpart and the total number of particles to be classified, is reasonably high for every algorithm, which evidences the success of our quantum implementations.
%\begin{table}[th!]
\begin{longtable}{| p{0.45cm} | p{1.6cm} | p{2.8cm} | p{2.0cm} | p{2.6cm} | p{3.1cm} | }
    %\begin{center}
\hline
& \centering{ Quantum \texttt{K-means}} & \centering {Quantum Affinity Propagation}  & \centering {Quantum $k_T$, $p=1$} & \centering{Quantum anti-$k_T$, $p=-1$}  & \centering {Quantum Cam/Aachen, $p=0$}   \cr   \hline 
 \centering $\varepsilon_c$ &\centering 0.94 &\centering 1.00 &\centering 0.98 &\centering 0.99 &\centering 0.98 \cr   \hline

\omit
   \\     
\caption{Efficiencies of the different quantum algorithms in comparison with their classical counterparts.}
\label{tab:eff}
    %\end{center}
\end{longtable}
%\end{table}
\section{Conclusions}

We have considered the quantum versions of the well-known \texttt{K-means}, \texttt{AP} and $k_T$-jet clustering algorithms.
These quantum versions, which are based on two novel quantum procedures, might speed up their classical counterparts  on a quantum device with qRAM. The first one is a quantum subroutine which serves to compute distances satisfying Minkowski metric, whereas the second one consists of a quantum circuit to track the rough maximum into a list of unsorted data.

In the case of the \texttt{K-means} clustering algorithm, applying both quantum subroutines we would get an exponential advantage in the vector dimensionality $d$, as well as in the number of clusters $K$.
In the  quantum version of the Affinity Propagation method the similarity is computed with the same quantum procedure as in the \texttt{K-means} case. Thus, it would lead to an exponential speed-up regarding its classical counterpart in the vector dimensionality $d$.
Finally, we have presented the quantum versions of the well-known $k_T$-jet clustering algorithms. On a true universal quantum device, the implementation of these algorithms would exhibit an exponential speed up in finding the minimum distance.
In this way, we would obtain a quantum version of order $\mathcal{O}(N\log(N))$, which is of the same order as the fully optimal version of \texttt{FastJet}.

For all the clustering algorithms considered, the quantum simulations show an excellent performance and clustering efficiencies. Furthermore, the comparison with their classical counterparts displays that both classifications of the LHC simulated data are quite in agreement.

%\acknowledgments
\noindent {\bf Acknowledgments}\\ 
\noindent This work is supported by the Spanish Government (Agencia Estatal de Investigaci\'on MCIN/AEI/ 10.13039/501100011033) Grant
No. PID2020-114473GB-I00, and Generalitat Valenciana
Grant No. PROMETEO/2021/071. JJML is supported by Universitat de Val\`encia, Atracci\'o de Talent programme. LC is supported by Generalitat Valenciana GenT Excellence Programme (CIDEGENT/ 2020/011).

%\begin{thebibliography}{99}
%\bibitem{...}
%....
\begin{normalsize}
\bibliography{bibliography}

\providecommand{\href}[2]{#2}\begingroup\raggedright\begin{thebibliography}{10}

\bibitem{Wei:2019rqy}
A.Y.~Wei, P.~Naik, A.W.~Harrow and J.~Thaler, \emph{{Quantum Algorithms for Jet
  Clustering}}, \href{https://doi.org/10.1103/PhysRevD.101.094015}{\emph{Phys.
  Rev. D} {\bfseries 101} (2020) 094015}
  [\href{https://arxiv.org/abs/1908.08949}{{\ttfamily 1908.08949}}].

\bibitem{Pires:2021fka}
D.~Pires, P.~Bargassa, J.a.~Seixas and Y.~Omar, \emph{{A Digital Quantum
  Algorithm for Jet Clustering in High-Energy Physics}},
  \href{https://arxiv.org/abs/2101.05618}{{\ttfamily 2101.05618}}.

\bibitem{Delgado:2022snu}
A.~Delgado and J.~Thaler, \emph{{Quantum Annealing for Jet Clustering with
  Thrust}},  \href{https://arxiv.org/abs/2205.02814}{{\ttfamily 2205.02814}}.

\bibitem{deLejarza:2022bwc}
J.J.M.~de~Lejarza, L.~Cieri and G.~Rodrigo, \emph{{Quantum clustering and jet
  reconstruction at the LHC}},
  \href{https://doi.org/10.1103/PhysRevD.106.036021}{\emph{Phys. Rev. D}
  {\bfseries 106} (2022) 036021}
  [\href{https://arxiv.org/abs/2204.06496}{{\ttfamily 2204.06496}}].

\bibitem{Ramirez-Uribe:2021ubp}
S.~Ram\'\i{}rez-Uribe, A.E.~Renter\'\i{}a-Olivo, G.~Rodrigo, G.F.R.~Sborlini
  and L.~Vale~Silva, \emph{{Quantum algorithm for Feynman loop integrals}},
  \href{https://doi.org/10.1007/JHEP05(2022)100}{\emph{JHEP} {\bfseries 05}
  (2022) 100} [\href{https://arxiv.org/abs/2105.08703}{{\ttfamily
  2105.08703}}].

\bibitem{Delgado:2022tpc}
A.~Delgado et~al., \emph{{Quantum Computing for Data Analysis in High-Energy
  Physics}},  in \emph{{2022 Snowmass Summer Study}}, 3, 2022
  [\href{https://arxiv.org/abs/2203.08805}{{\ttfamily 2203.08805}}].

\bibitem{macqueen1967some}
J.~MacQueen et~al., \emph{Some methods for classification and analysis of
  multivariate observations},  in \emph{Proceedings of the fifth Berkeley
  symposium on mathematical statistics and probability}, Oakland, CA, USA,
  1967.

\bibitem{Frey2007ClusteringBP}
B.J.~Frey and D.~Dueck, \emph{Clustering by passing messages between data
  points}, \href{https://doi.org/10.1126/science.1136800}{\emph{Science}
  {\bfseries 315} (2007) 972}.

\bibitem{Catani:1991hj}
S.~Catani, Y.L.~Dokshitzer, M.~Olsson, G.~Turnock and B.R.~Webber, \emph{{New
  clustering algorithm for multi - jet cross-sections in e+ e- annihilation}},
  \href{https://doi.org/10.1016/0370-2693(91)90196-W}{\emph{Phys. Lett. B}
  {\bfseries 269} (1991) 432}.

\bibitem{Catani:1993hr}
S.~Catani, Y.L.~Dokshitzer, M.H.~Seymour and B.R.~Webber, \emph{{Longitudinally
  invariant $K_t$ clustering algorithms for hadron hadron collisions}},
  \href{https://doi.org/10.1016/0550-3213(93)90166-M}{\emph{Nucl. Phys. B}
  {\bfseries 406} (1993) 187}.

\bibitem{Ellis:1993tq}
S.D.~Ellis and D.E.~Soper, \emph{{Successive combination jet algorithm for
  hadron collisions}},
  \href{https://doi.org/10.1103/PhysRevD.48.3160}{\emph{Phys. Rev. D}
  {\bfseries 48} (1993) 3160}
  [\href{https://arxiv.org/abs/hep-ph/9305266}{{\ttfamily hep-ph/9305266}}].

\bibitem{Cacciari:2008gp}
M.~Cacciari, G.P.~Salam and G.~Soyez, \emph{{The anti-$k_t$ jet clustering
  algorithm}}, \href{https://doi.org/10.1088/1126-6708/2008/04/063}{\emph{JHEP}
  {\bfseries 04} (2008) 063} [\href{https://arxiv.org/abs/0802.1189}{{\ttfamily
  0802.1189}}].

\bibitem{Buhrman:2001}
H.~Buhrman, R.~Cleve, J.~Watrous and R.~de~Wolf, \emph{Quantum fingerprinting},
  \href{https://doi.org/10.1103/PhysRevLett.87.167902}{\emph{Phys. Rev. Lett.}
  {\bfseries 87} (2001) 167902}.

\bibitem{Grover:1997fa}
L.K.~Grover, \emph{{Quantum mechanics helps in searching for a needle in a
  haystack}}, \href{https://doi.org/10.1103/PhysRevLett.79.325}{\emph{Phys.
  Rev. Lett.} {\bfseries 79} (1997) 325}
  [\href{https://arxiv.org/abs/quant-ph/9706033}{{\ttfamily
  quant-ph/9706033}}].

\end{thebibliography}\endgroup
\end{normalsize}

%\end{thebibliography}

\end{document}